\documentclass[aps,prb,groupedaddress,twocolumn]{revtex4}

\usepackage{graphicx}

\begin{document}

\title{Detection of weak-order phase transitions in ferromagnets by ac resistometry}

\author{V.~V.~Khovailo}
\author{T.~Abe}

\affiliation{National Institute of Advanced Industrial Science and
Technology, Tohoku Center, Sendai 983--8551, Japan}

\author{T.~Takagi}

\affiliation{Institute of Fluid Science, Tohoku University, Sendai
980--8577, Japan}

\begin{abstract}
It is shown that ac resistometry can serve as an effective tool
for the detection of phase transitions, such as spin reorientation
or premartensitic phase transitions, which generally are not
disclosed by dc resistivity measurement. Measurement of
temperature dependence of impedance, $Z(T)$, allows one to unmask
the anomaly, corresponding to a weak-order phase transition. The
appearance of such an anomaly is accounted for by a change in the
effective permeability $\mu$ of a sample upon the phase
transition. Moreover, frequency dependence of $\mu$ makes it
possible to use the frequency of the applied ac current as an
adjusting parameter in order to make this anomaly more pronounced.
The applicability of this method is tested for the rare earth Gd
and Heusler alloy Ni$_2$MnGa.
\end{abstract}

\maketitle

It is well known that impedance $Z$ of a magnetic conductor
depends on the effective permeability $\mu$ of the material. In
particular, this is evident from the fact that impedance of soft
ferromagnetic materials changes considerably in a magnetic field,
which is accounted for by a drastic change of $\mu$ upon
application of the field. This phenomenon, called giant
magnetoimpedance, is most pronounced in soft ferromagnetic wires
with a peculiar configuration of magnetic domains (Refs.~1-3 and
references therein), but generally it seems to be an intrinsic
property of soft ferromagnets and recently it was observed in
several Heusler alloys.~\cite{4-k,5-f}

The fact that $Z$ depends on the effective permeability $\mu$
gives rise to suggest that it can be used for studying weak-order
phase transitions, such as spin reorientation, order-disorder, and
so on, occurring in the ferromagnetic state. Indeed, since dc
resistivity does not allow detection or accurate determination of
a temperature of the transition, upon which no significant change
in the Fermi surface, mean free path or electrons concentration
occurs, such transitions are usually studied by magnetic
measurements. The existence of the anomaly corresponding to a
weak-order phase transition at a temperature dependence of
magnetization evidences that such a transition is accompanied by a
change in permeability of the sample, therefore, measurements of
$Z(T)$ could also be effectively used for this aim. Moreover, in
certain cases, this method can have an advantage compared to the
magnetic measurements, for example, if a weak-order phase
transition is located in the proximity to another phase
transition, characterized by a drastic change in the
magnetization.

It is shown in this article, that ac resistometry can serve as an
effective tool for this purpose. The validity of this method is
demonstrated for the case of the rare earth Gd, which undergoes a
spin reorientation transition at a low temperature, and for a
Ni$_2$MnGa alloy, which undergoes a weak-order premartensitic
phase transition.

Impedance $Z$ of a cylindrical conductor can be derived using
Maxwell equations:~\cite{6-l}

$$ Z = R+iX = \frac{1}{2}R_{dc} ka \frac{J_0(ka)}{J_1(ka)}, $$

\noindent where $R_{dc}$ is dc resistance of the conductor at a
frequency $f = \omega /2\pi = 0$, $a$ is the conductor radius,
$J_i$ are Bessel functions of the first kind, and $k =
(1+i)/\delta$, where $\delta$ is the skin depth,

$$ \delta = c\sqrt{\frac{\rho}{2\pi \omega \mu}} . $$

\noindent The skin depth $\delta$ is determined by resistivity
$\rho$ and effective permeability $\mu$ of the conductor, and by
the frequency $\omega$ of applied to the conductor ac current. In
the case of a slab geometry, the expression for the impedance
is~\cite{7-m}

$$ Z = R_{dc}ikd \: \mathrm{coth}(ikd) ,  $$

\noindent where $d$ is the half thickness. Both of these equations
show that at a constant frequency the temperature dependence of
$Z$ is determined by the temperature dependence of $\rho$ and
$\mu$ as $Z(T) \sim \sqrt{\rho (T)\mu (T)}$.

The experimental setup for the measurement of temperature
dependencies of the impedance $Z$ consists of a lock-in amplifier,
a function synthesizer, a reference resistor connected in a serial
way with the sample, and a computer. Alternating voltage of
amplitude $V_0$ at a frequency $f$ was applied to the reference
resistor using the function synthesizer. The voltage drop on the
sample and the phase shift between the voltage and the reference
signal were measured by the lock-in amplifier.

Measurements of $\rho (T)$ and $Z(T)$ were performed on a Gd
sample measuring $8 \times 2 \times 1$ mm$^3$ and a Ni$_2$MnGa
sample measuring $10 \times 1.5 \times 0.5$ mm$^3$. Using typical
values of $\rho$ and $\mu$ for Gd, $\rho \approx 130 \times
10^{-6}$ $\Omega \: \mathrm{cm}$ and $\mu \approx 200$, and
Ni$_2$MnGa, $\rho \approx 30 \times 10^{-6}$ $\Omega \:
\mathrm{cm}$ and $\mu \approx 200$ (Ref.~4), the skin depth for
$7~\mathrm{kHz} < f < 100~\mathrm{kHz}$ is $1500~\mathrm{\AA} <
\delta < 5000~\mathrm{\AA}$  for the Gd sample and
$700~\mathrm{\AA} < \delta < 2600~\mathrm{\AA}$ for the Ni$_2$MnGa
sample. These values are much smaller than typical sample
thickness ($\ge 0.5$ mm), so that the skin effect is essential at
these frequencies.

\medskip

\textbf{A. Spin reorientation transition in Gd}

It is known from the literature that Gd orders ferromagnetically
at Curie temperature $T_C \approx 293$~K and undergoes a spin
reorientation transition at a lower temperature, $T_s \approx
225$~K (Refs. 8 and 9 and references therein).

\begin{figure}[t]
\begin{center}
\includegraphics[width=7cm]{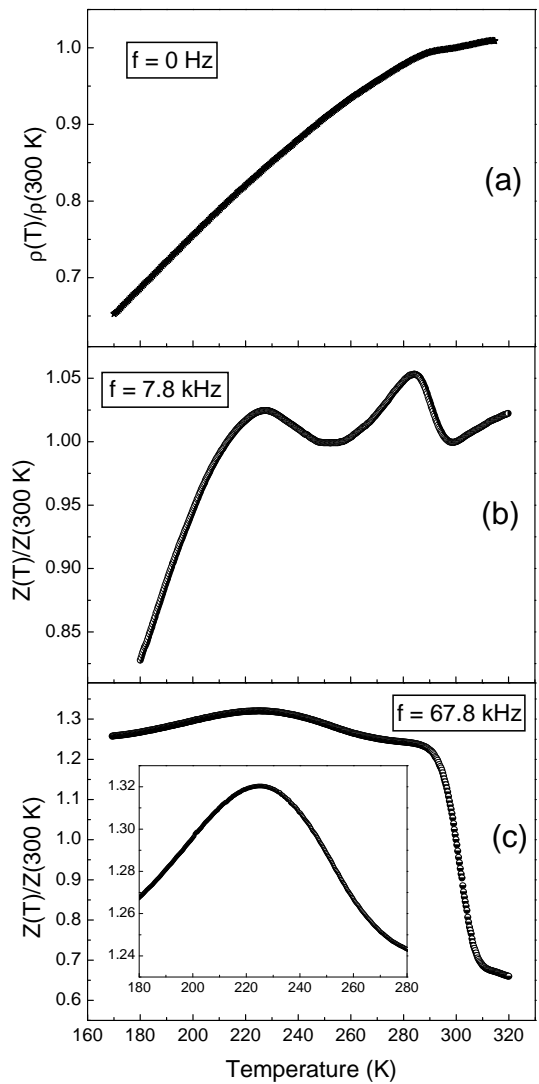}
\caption{Temperature dependencies of resistivity $\rho$ and
impedance $Z$ measured on the Gd sample.}
\end{center}
\end{figure}

\begin{figure}[th]
\begin{center}
\includegraphics[width=7cm]{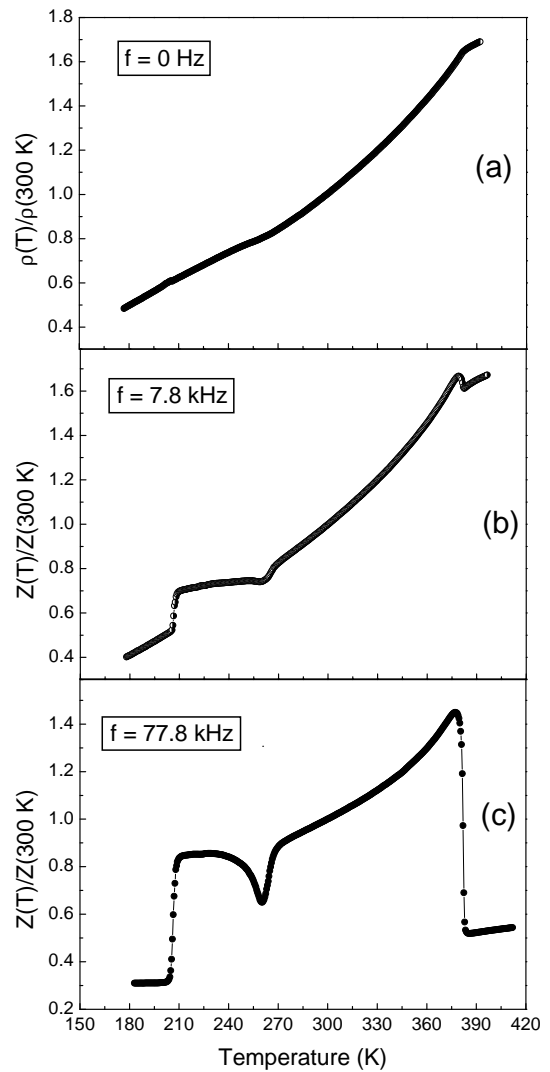}
\caption{Temperature dependencies of resistivity $\rho$ and
impedance $Z$ measured on the Ni$_2$MnGa sample.}
\end{center}
\end{figure}

Temperature dependencies of $\rho$ and $Z$, measured upon heating,
are presented in Fig.~1. As evident from Fig.~1(a), $\rho (T)$ has
an anomaly at Curie temperature $T_C$, equal to 290~K for this
sample. No anomaly, corresponding to the spin reorientation
transition is seen on the curve. This result is in agreement with
early transport measurements of Gd.~\cite{9a-n}

Temperature dependencies of the impedance $Z$, measured on this
sample [Figs.~1(b) and 1(c)], drastically differ from the
temperature dependence of the resistivity $\rho$. $Z(T)$, measured
at a frequency $f = 7.8$~kHz has a complex temperature dependence
[Fig.~1(b)], namely, two marked peaks are clearly seen on this
curve. The first peak is observed at the temperature of spin
reorientation phase transition, $T_s = 226$~K. This temperature,
determined from the $Z(T)$ measurement, is in good agreement with
the temperature determined by another experimental
techniques.~\cite{8-d,9-c}  Since $\rho (T)$ does not exhibit
anomaly at $T_s$, the local maximum of $Z(T)$ at this temperature
is accounted for by a larger permeability $\mu$ at the spin
reorientation transition temperature. The origin of the second
peak observed near the ferromagnetic phase transition is
presumably the same because Gd has a pronounced peak of
susceptibility $\chi = (\mu - 1)/4\pi$ in the vicinity of $T_C$
(Ref.~9). The result of $Z(T)$ measurement performed at a higher
frequency, $f = 67.8$~kHz, is shown in Fig.~1(c). This curve also
has a pronounced peak at the spin reorientation temperature $T_s$.
Since upon transition from the ferromagnetic to paramagnetic state
the impedance $Z$ drastically decreases, this peak is shown in the
inset in Fig.~1(c). The existence of this prominent and
well-defined peak permits an accurate determination of the spin
reorientation transition temperature from the $Z(T)$ measurements.

Therefore, these results have shown that measurement of impedance
$Z$ can be effectively used, along with magnetic measurements, for
study spin reorientation transitions and construction of phase
diagrams of compounds undergoing such transitions.

\medskip

\textbf{B. Premartensitic transition in Ni$_2$MnGa}

An interesting property of near stoichiometric Ni$_2$MnGa alloys
is that they undergo a so-called premartensitic phase transition,
which is characterized by a modulation of the Heusler cubic
structure.~\cite{10-z} Resistivity measurement of these alloys
showed~\cite{11-k,12-c,13-z} that it is difficult to determine the
temperature of premartensitic and martensitic phase transition
from $\rho (T)$ data because the resistivity has weak and broad
anomalies at these phase transformation temperatures. Measurement
of magnetization is an effective tool to determine martensitic
transformation temperature,~\cite{14-w} but as for the
premartensitic transition, $M(T)$ measurement is sometimes not
sufficient to determine the temperature of the premartensitic
transition in the case of polycrystalline samples.~\cite{15-z} In
order to proove that the premartensitic transition can be
disclosed by $Z(T)$ measurements, temperature dependencies of
$\rho$ and $Z$ were measured on a polycrystalline sample of
stoichiometric Ni$_2$MnGa composition.

Measurement of $\rho (T)$, shown in Fig.~2(a), revealed typical
for the stoichiometric Ni$_2$MnGa behavior of
resistivity.~\cite{11-k,12-c,13-z} A marked anomaly is seen only
at the ferromagnetic transition temperature $T_C = 380$~K, whereas
the martensitic and premartensitic phase transitions are
accompanied by a broad change in the slope of the curve at $T_m
\approx 200$~K and $T_P \approx 260$~K, respectively.

Unlike the resistivity $\rho$, the impedance $Z$ of this sample
has quite different temperature dependence [Figs.~2(b) and 2(c)].
$Z(T)$ measured at $f = 7.8$~kHz [Fig.~2(b)] exhibits a peak in
the vicinity of Curie temperature $T_C$, which is accounted for by
a high susceptibility of the sample. Contrary to $\rho (T)$, the
temperature dependence of the impedance $Z$ exhibits a jump-like
behavior at the martensitic transition temperature $T_m$. This is
due to the fact that the martensitic phase has a lower
permeability $\mu$ as compare to the austenitic cubic phase.
Finally, the premartensitic transition appears at this curve as a
small dip at $T_P = 260$~K. Measurement of $Z(T)$ at a higher
frequency $f = 77.8$~kHz [Fig.~2(c)] indicates that the anomalies
corresponding to the phase transitions are enhanced. The most
interesting finding is that the increase in the frequency of the
current results in the appearance of a pronounced dip at the
premartensitic phase transition. This observation allows an
accurate determination of the premartensitic transition
temperature $T_P$. Since the impedance $Z$ decreases and then
increases smoothly around $T_P$, whereas the drastic drop of $Z$
is observed at $T_m$, it can be suggested that by measuring $Z(T)$
one can easy distinguish these transitions even if they are close
to each other.

\smallskip

In conclusion, the results presented in this article have shown
that, by measuring the temperature dependence of impedance $Z$, it
is possible to unmask anomalies which are generally not observed
on dc resistivity curves. This has been confirmed for the case of
gadolinium, which exhibits a spin reorientation transition at $T_s
= 226$~K, and for the case of Ni$_2$MnGa, which exhibits a
premartensitic phase transition at $T_P = 260$~K. Moreover, the
results of this study indicated that by adjusting the frequency of
the ac current one can observe a sharp anomaly at the temperature
of such a transition. Therefore, this method can be used, along
with magnetic measurements, as a simple and effective tool for the
study of spin reorientation transitions and construction of phase
diagrams in intensively studied rare-earth alloys and in other
magnetic alloys and compounds.

\medskip

The authors are thankful to Professor R.~Z.~Levitin for providing
the gadolinium sample. A Postdoctoral Fellowship Award from the
Japan Society for the Promotion of Science (JSPS) is greatly
acknowledged.

\end{document}